\documentclass[iop]{emulateapj}
\usepackage{apjfonts}
\usepackage{graphicx}
\usepackage{color}

\usepackage{epstopdf}
\usepackage{amsmath}

\def\ltsima{$\; \buildrel < \over \sim \;$}
\def\simlt{\lower.5ex\hbox{\ltsima}}
\def\gtsima{$\; \buildrel > \over \sim \;$}
\def\simgt{\lower.5ex\hbox{\gtsima}}

\newcommand\lsim{\mathrel{\rlap{\lower4pt\hbox{\hskip1pt$\sim$}}
\raise1pt\hbox{$<$}}}
\newcommand\gsim{\mathrel{\rlap{\lower4pt\hbox{\hskip1pt$\sim$}}
\raise1pt\hbox{$>$}}}

\allowdisplaybreaks
\usepackage{color}

\shorttitle{Sgr A* BFF  }
\shortauthors{Naoz et al. }

\begin{document}

\title{  A hidden friend for the galactic center black hole, Sgr A*}

\author{  Smadar Naoz$^{1,2}$,  Clifford M. Will$^{3,4}$, Enrico Ramirez-Ruiz$^{5,6}$, Aur\'elien Hees$^7$,  Andrea M.~Ghez$^1$, Tuan Do$^1$}

\altaffiltext{1}{Department of Physics and Astronomy, University of California, Los Angeles, CA 90095, USA}
\altaffiltext{2}{Mani L. Bhaumik Institute for Theoretical Physics, Department of Physics and Astronomy, UCLA, Los Angeles, CA 90095, USA}
\altaffiltext{3}{Department of Physics, University of Florida, Gainesville, FL 32611, USA}
\altaffiltext{4}{Institut d'Astrophysique, Sorbonne Universit\'e, 75014 Paris, France}
\altaffiltext{5}{Department  of  Astronomy  and  Astrophysics,  University  of
California, Santa Cruz, CA 95064,USA}
\altaffiltext{6}{Niels Bohr Institute, University of Copenhagen, Blegdamsvej 17, 2100 Copenhagen, Denmark}
\altaffiltext{7}{SYRTE, Observatoire de Paris, Universit\'e PSL, CNRS, Sorbonne Universit\'e, LNE, 61 avenue de l'Observatoire, F-75014 Paris, France}


\begin{abstract}

The hierarchical nature of galaxy formation suggests that a supermassive black hole binary could exist in our galactic center. We propose a new approach to constraining  the possible orbital configuration of such a binary companion to the galactic center black hole Sgr A*  through the measurement of stellar orbits. 
 Focusing  on the star S0-2,  we show that requiring its {\it orbital stability} in the presence of a companion to Sgr A* yields stringent constraints on the possible configurations of such a companion. Furthermore, we show that precise measurements of {\it time variations} in the orbital parameters of S0-2  could yield stronger constraints.  Using existing data on S0-2 we derive upper limits on the  binary black hole separation as a function of the companion mass. For the case of a circular orbit, we can rule out a $10^5 \,M_\odot$ companion  with a semimajor axis greater than 170 astronomical units or $0.8$ mpc. 
 This is already more stringent than bounds obtained from studies of the proper motion of Sgr A*. Including other stars orbiting  the galactic center should yield stronger constraints that could help uncover the presence of a companion to Sgr A*.  We show that a companion can also affect the accretion process, resulting in a variability which may be consistent with the measured infrared flaring timescales and amplitudes.  Finally, if such a companion exists, it will emit gravitational wave radiation, potentially detectable with LISA. 

\end{abstract}


\maketitle

\section{INTRODUCTION}

Almost every galaxy, our own Milky Way included, harbors a supermassive black hole (SMBH) in its nucleus.  Furthermore, the hierarchical nature of the galaxy formation paradigm suggests that major galaxy mergers may result in the formation of {\em binary} SMBHs \citep[e.g.,][]{DiMatteo+05,Hopkins+06,Robertson+06,Callegari+09}. Already observations have suggested  several   
{\it wide} binary systems as well as binary candidates with sub-parsec  to tens to hundreds of parsec separations 
\citep[e.g.,][]{Sillanpaa+88,Rodriguez+06,Komossa+08,Bogdanovic+09,Boroson+09,Dotti+09,Batcheldor+10,Deane+14,Liu+14,Liu+17,Li+16,Bansal+17,Kharb+17,Runnoe+17,Pesce+18}.   Furthermore,  observations of several active galactic nuclei pairs with kpc-scale  separations have been interpreted as systems containing SMBH binaries \citep[e.g.,][]{Komossa+03,Bianchi+08,Comerford+09bin,Liu+10kpc,Green+10,Smith+10,Comerford+18}.

 The most definitive case of the existence of an SMBH is at the center of our galaxy, commonly known as Sagittarius A* (Sgr A*).
Recent advances in technology, such as the advent of adaptive optics (AO), have made it possible to  measure stellar orbits at the galactic center. These orbits imply the presence of a 4 million solar masses black hole 
\citep[e.g.,][]{Ghez+00,Ghez+08,Gillessen+09,boehle:2016wu,gillessen:2017aa,gravity:2019aa}, residing in a 
 dense stellar environment, called a nuclear star cluster \citep[e.g.,][]{Ghez+03,Gillessen+09,Lu+13}.  Continued observations have enabled precision measurements of the distance to the galactic center \citep[e.g..][]{Ghez+03,boehle:2016wu,gravity:2019aa}, new constrains on the fifth force \citep{HEss+17} and the first gravitational redshift measurements near a SMBH \citep{GRAVITY18,Do+19}.

 Using measurements of stellar orbits,  we  examine here the possibility that the black hole at the center of our galaxy has a companion.
If Sgr A* has a hidden companion, a myriad of observable effects may occur. 
First we consider the allowable binary configuration under the indirect assumption that S0-2 is stable against three body scattering (Sec.~\ref{sec:analytical}). We then focus on three direct observational signatures (Sec.~\ref{sec:obs}).
First, the varying gravitational field of the binary  could induce observable perturbations on the orbits of S0-2 and other stars surrounding the pair (Sections \ref{sec:stability} and \ref{sec:S02dynamics}).
Second, if there is a disk of accreting matter surrounding the massive member of the pair, the passage of the secondary black hole through the disk could induce variability in the output of electromagnetic radiation from the vicinity of Sgr A*,  the emissive source associated with the SMBH (Sec.~\ref{sec:IRvir}). Finally, such a companion may be detected by the future LISA gravitational-wave mission (Sec.~\ref{sec:GW}).
Current and future observations could then either reveal the presence of such a companion or provide constraints on its mass and orbital parameters.

\begin{figure}
  \begin{center}
    \includegraphics[width=\linewidth]{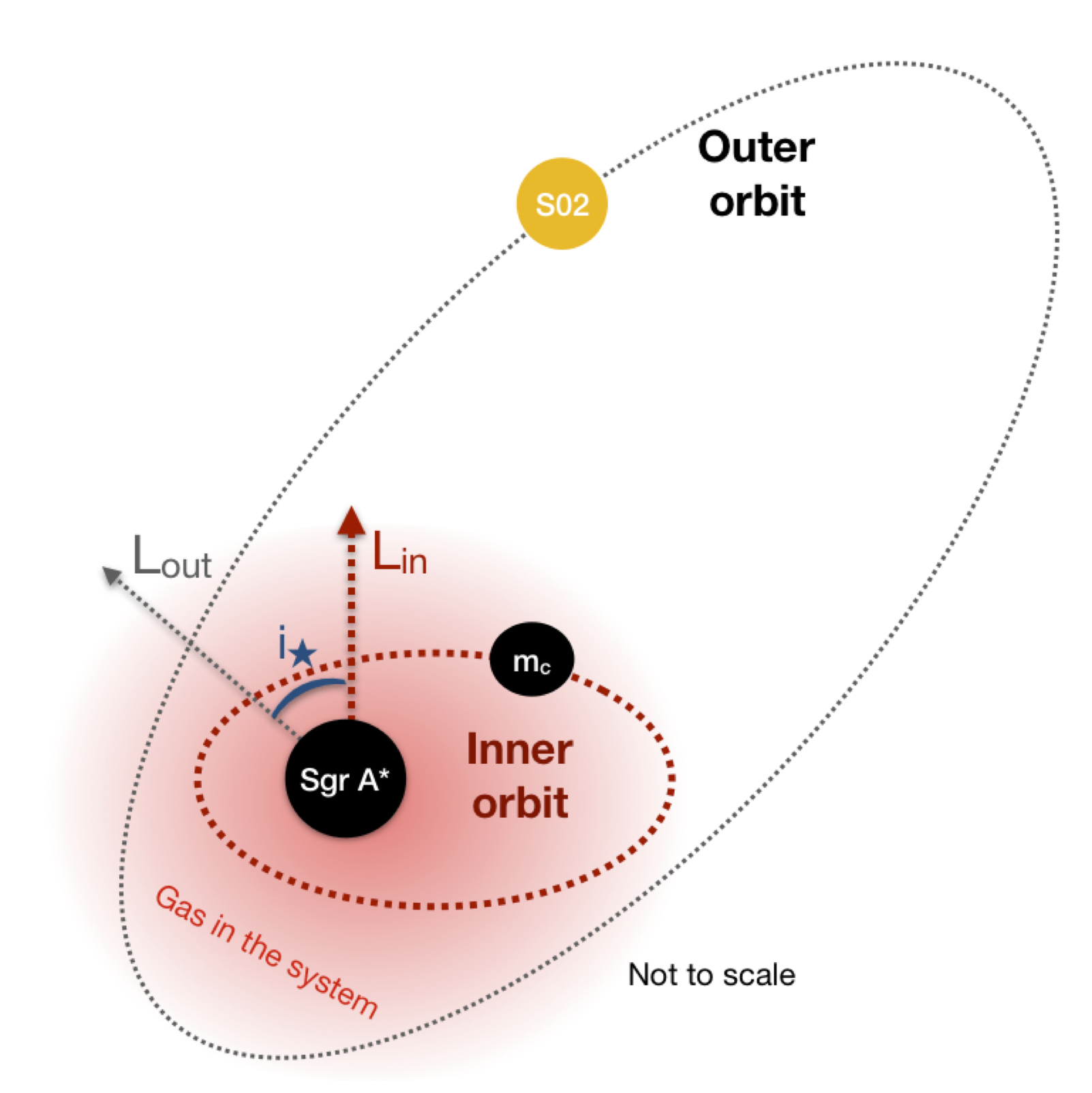}
  \end{center} 
  \caption{  \upshape A hierarchical three-body system consisting of the inner binary and the star S0-2.   } \label{fig:Cartoon} 
\end{figure}

\section{Stellar perturbations from an inner binary system  }\label{sec:analytical}

We consider a hierarchical triple system in which the inner binary consists of the massive black hole Sgr A* and a lighter black-hole companion, with masses $m_{\bullet}$ and $m_c$, respectively, and the outer body is a star of mass $m_\star$ such as S0-2
(see  Fig.\ \ref{fig:Cartoon}).
We assume that the ratio $a_c/a_\star$ of the inner and outer semimajor axes is small, or that the inner orbital period is short compared to the outer orbital period.  But unlike conventional hierarchical triple systems, where the outer body perturbs the inner binary, inducing Eccentric Kozai-Lidov oscillations, for example, here we treat the outer body as a massless test particle.  It has no effect on the inner binary, but its orbit is perturbed by the varying multipole moments of the inner binary's gravitational field.  We call these ``inverse Eccentric Kozai-Lidov'' (iEKL) perturbations \citep[e.g.,][]{Naoz+17,Zanardi+17}.  For simplicity we assume that the two black holes have zero spin.  

Such outer test particle systems were previously studied to the quadrupole level of approximation \citep[e.g.,][]{Ziglin75,Gallardo+12,Verrier+09,Farago+10}, and were recently extended to octupole order \citep{Naoz+17,Zanardi+17} and to hexadecapole order \citep{Vinson+18,DeElia+19}.  We choose a fixed reference coordinate system (invariable plane) whose $Z$-axis is parallel to the system's angular momentum vector  \citep[see e.g.,][]{Naoz+11sec}, which in this case comes entirely from the inner binary.  The orbit of the third, outer body is described by inclination $i_\star$ of its plane relative to the $XY$-plane, angle of ascending node $\Omega_\star$ relative to the $X$-axis, and pericenter angle $\omega_\star$ relative to the line of nodes.  The orbit itself is characterized by semimajor axis $a_\star$ and eccentricity $e_\star$.  The inner orbit has semimajor axis $a_c$ and eccentricity $e_c$.  However, its inclination vanishes, and as a result, in the absence of an outer body, its nodal and pericenter angles  $\Omega_c$ and $\omega_c$ would be ambiguous; only the sum $\varpi_c \equiv \omega_c + \Omega_c$, the angle from the $X$-axis to the pericenter, is well defined.  However, the outer body's orbit defines a nodal angle $\Omega_\star$, and in the case of non-zero masses for all three bodies, it is known that $\Omega_c = \Omega_\star + \pi$.  Accordingly, we will retain this definition in the test-mass outer body limit.  This then serves to {\em define} the inner orbit's pericenter angle $\omega_c \equiv \varpi_c - \Omega_\star - \pi$.  This will be important when we introduce general relativistic effects, which induce a precession of $\varpi_c$.  

These angles are defined in the invariable plane and should not be confused with the observed inclination, ascending node and pericenter angles ($i_{\star\rm ,sky}$ $\Omega_{\star \rm ,sky}$, $\omega_{\star \rm ,sky}$), defined with respect to the line of sight (see Section 3.1.1 from \cite{ghez:2005dq}).  Given a configuration of the inner SMBH binary, simple relations between these angles can be obtained.  
Here we work in the reference frame of the invariable plane. To indicate parameters on the plane of the sky we will add the subscript ``sky.''

For the outer orbit, at quadrupole order and averaging over both the inner and outer orbital periods (the ``secular approximation''), $a_\star$ and $e_\star$ are constant. For our analysis, we consider the time evolution of $\Omega_\star$, $\theta = \cos i_\star$, and the variable $\varpi_\star$, defined by the relation $d\varpi_\star/dt \equiv d\omega_\star/dt + \cos i_\star \, d\Omega_\star/dt$, \citep[the full set of equations can be found in][]{Naoz+17}
\begin{align}
\frac{d\Omega_\star}{d\tau} &=-\frac{3\pi}{4} \eta \alpha^2 \frac{\theta Q}{(1-e_\star^2)^2} \,,
\label{eq:bigOmega}
\\
\frac{d\theta}{d\tau} &=\frac{15\pi}{4} \eta \alpha^2  \frac{e_c^2(1-\theta^2)\sin 2 \omega_c}{(1-e_\star^2)^2} \,,
\label{eq:theta}
\\
\frac{d\varpi_\star}{d\tau}&=\frac{3\pi}{8} \eta \alpha^2 \frac{4+6e_c^2 - 3(1-\theta^2)Q}{(1-e_\star^2)^2} \,,
\label{eq:varpi}
\end{align}
where $\eta \equiv m_{\bullet} m_c/(m_\bullet+m_c)^2$, $\alpha \equiv a_c/a_\star$, and $Q\equiv 2+3e_c^2-5e_c^2 \cos 2\omega_c$. The parameter $\tau$ is time measured in units of the outer orbital period. 
The timescale for these quadrupolar precessions is given, from Eqs.\ (\ref{eq:bigOmega}) and (\ref{eq:theta}) by \citep{Naoz+17}: 
\begin{equation}\label{eq:quadrupole}
t_{\rm quad}  \sim \frac{4}{3} P_{\star} (1-e_{\star}^2)^2 \frac{ m^2}{m_{\bullet} m_c}  \left( \frac{a_{\star}}{a_c} \right)^2  \ , 
\end{equation}
where $m \equiv m_\bullet +m_c$ and $P_{\star}$ is the orbital period of the star.
Note that this timescale is different from the nominal EKL timescale \citep[see][for further discussion]{Naoz16,Naoz+17}. 

 Earlier studies \citep[e.g.,][]{Naoz+17,DeElia+19}, showed that octupole terms in the secular outer test particle equations (i.e., iEKL) can lead to large eccentricity excitations of the outer orbit. These large eccentricity values can lead to instability by plunging the star so close to the SMBH binary that it experiences a three-body scattering event. To avoid this we require that general relativistic (GR) precessions be strong enough to suppress quadrupole, and thus octupole, excitations \citep[e.g.,][]{Naoz+12GR,Naoz+17}.  We include the octupole level of approximation in the numerical results presented in Appendix \ref{sec:numeric}.

The leading GR pericenter precession effects on the orbits are given by 
\begin{align}
\left (\frac{d\varpi_c}{dt}\right )_{GR} &= \frac{6\pi Gm}{P_{c} c^2 a_c(1-e_c^2)} \,, 
\label{eq:grprecessin} \\
\left (\frac{d\varpi_\star}{dt} \right )_{GR} &= \frac{6\pi Gm}{P_{\star} c^2 a_\star(1-e_\star^2)} \,,
\label{eq:grprecessout}
\end{align}
where  $G$ and $c$ are Newton's constant and the speed of light.
We also consider the effects of gravitational-wave (GW) damping on the inner orbit. The timescale associated with this damping  is estimated as:
 \begin{eqnarray}\label{eq:grdamping}
 t_{GW} &\sim& 2\times 10^{8}~{\rm yr} \left(\frac{4\times 10^6~{\rm M}_\odot}{m_{\bullet}}\right)\left(\frac{10^4~{\rm M}_\odot}{m_{c}} \right) \nonumber \\ &\times& \left(\frac{m}{4\times 10^6}\right)^{-1} \left(\frac{a_c}{100 ~{\rm au}}\right)^4
 \end{eqnarray}
In appendix \ref{sec:numeric} we show an example of the numerical evolution of a SMBH binary orbited by a S0-2 like star (see Figure \ref{fig:example}).

 \begin{figure}[t]
     \begin{center}
     \includegraphics[width=\linewidth]{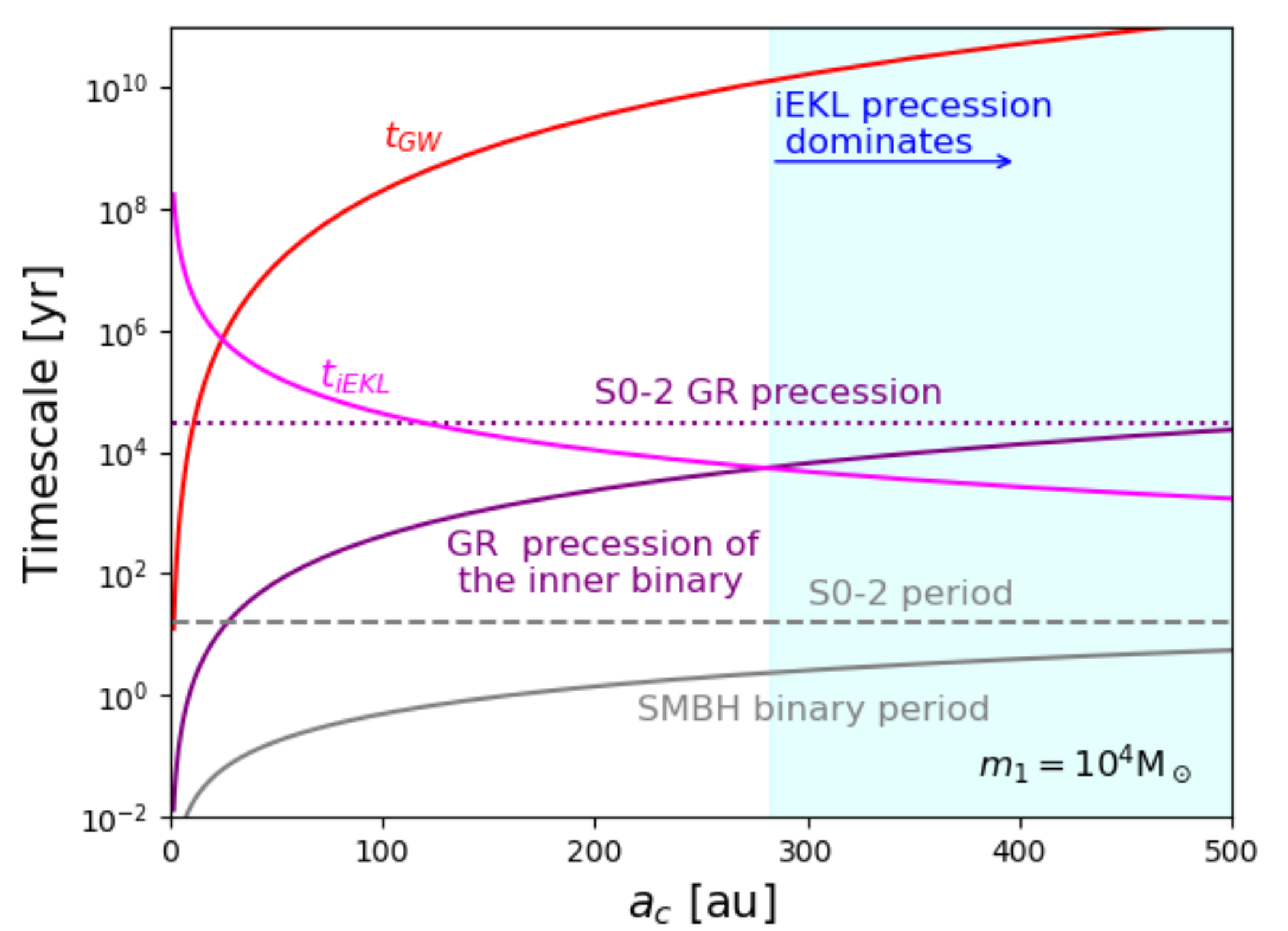}
     \end{center}
             \caption{ \upshape  Timescales of some of the different physical processes that affect the system. This example assumes a 
             $10^4 M_\odot$ companion to Sgr A*.  }
     \label{fig:timescales}
 \end{figure}

 \section{Observational Signatures and Constraints }\label{sec:obs}

  \subsection{Orbital Stability constraints}\label{sec:stability} 
 Around the black hole we observe the S-star cluster, whose presence implies a stable configuration. At the octupole level of approximation, the perturbations by a hypothetical inner binary can induce high eccentricity and thus scattering. Thus in order to avoid the destabilization of the cluster we derive one constraint on a companion by requiring that the  quadrupole and thus the octupole excitations are sufficiently suppressed.

At quadrupole order, the orbit is stable: its semimajor axis $a_\star$ and eccentricity $e_\star$ are constant, and only its orientation varies.   But when octupole-order terms are included, excitations of $e_\star$  can occur.
A rough rule of thumb for the quadrupole approximation to be valid is to require that \citep[e.g.,][]{Naoz+11sec}:
     \begin{equation}\label{eq:epsi}
 \epsilon= \eta \left(\frac{a_c}{a_{\star}}\right)\frac{e_c}{1-e_{\star}^2} < 0.1 \,,
\end{equation}
which merely encapsulates the requirement that the octupole perturbation be suitably small. 
However, this is only a rough criterion, and octupolar perturbations from the inner orbit may increase the outer test particle's eccentricity, possibly resulting in either an unbound orbit or in orbit crossing between the outer particle and the inner companion, leading to a scattering event. 
In fact, numerical simulations by \citet{DeElia+19} showed that the outer orbit eccentricity tends to grow beyond the values predicted by the secular approximation, which means that simply adopting Eq.\ (\ref{eq:epsi}) may be insufficient. Whether such ``instabilities'' are generic over the parameter space of interest or confined to very specific cases is a question that we will address separately. We note that we attempt to avoid this by requiring the stability regime at which GR precession is faster than the quadrupole timescale.

\begin{figure*}[!t]
  \begin{center} 
    \includegraphics[width=6in]{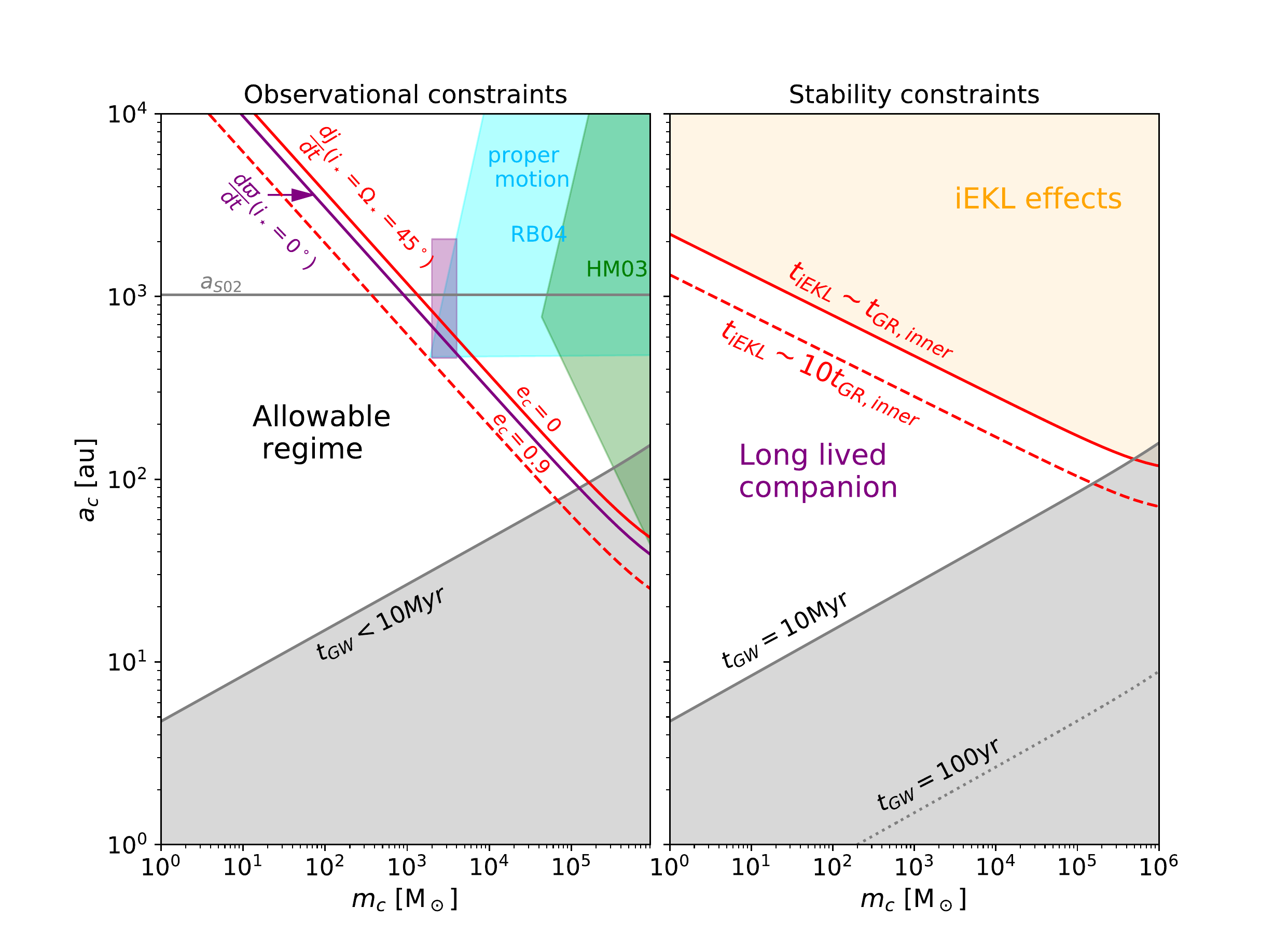}
  \end{center} 
  \caption{  \upshape {\bf Constraints on the mass-semimajor axis parameter space of a hypothetical companion to Sgr A*.}  {\it Right panel:}  We show constraints obtained by requiring that iEKL effects do not induce instabilities in the stellar orbit (yellow region).  Below the lines where the iEKL timescales are comparable to or 10 times the GR precession timescale, GR precessions tend to suppress iEKL excitations. The grey region denotes configurations where the GW merger time is shorter than 10 Myr (the dotted line is for a merger time of $100$yr.).  {\it Left panel:} We show observational  constraints obtained from the two invariant quantities of Eqs.~(\ref{eq:djdt})- (\ref{eq:dpomdtInv}). Specifically, we consider the bound on $d{\bf j}/dt$, for $i_\star=\Omega_{\star}=45^\circ$ and $e_c=0$ and $e_c=0.9$ (solid and dashed red lines, respectively). For $d\varpi/dt$, we show the bound on  the  square root of the sum of the squares of the two terms in Eq.~(\ref{eq:dpomdtInv}), assuming $i_\star=0^\circ$ and $e_c=0$. While   $i_\star$ is an unknown parameter, the two invariant variables have very different dependences on $i_\star$, enabling us to impose complementary bounds.   From these limits, {\it we can already rule out} a $10^5$~M$_\odot$ companion in a circular orbit beyond $170$ au. 
   We also include constraints imposed by data on the proper motion of Sgr A* \citep[][green]{Hansen+03}, and \citep[][cyan]{Reid+04}. We also show the exuded regime from   \citet{Gualandris+09} N-body integrations of stellar orbits in the presence of a companion, purple rectangle. 
    } \label{fig:allow} \vspace{0.2cm}
\end{figure*}

Since the resonant angle at the quadrupole level is $\omega_c$, \citep[e.g.,][]{Naoz+17,Zanardi+18,Vinson+18,DeElia+19} 
the general relativistic precession of the inner orbit may suppress Kozai-Lidov oscillations the orbit elements \citep[e.g.,][]{Naoz+12GR,Naoz+17}.
This occurs when the GR precession timescale is shorter that the iEKL timescale; comparing Eq.\  (\ref{eq:quadrupole}) with $d\varpi_c/dt$ from Eq.\ (\ref{eq:grprecessin}), we find the limiting 
separation of the companion,
 \begin{equation}\label{eq:limit}
a_{\rm lim} \sim  \left (\frac{4}{\eta} \right)^{2/9} R_g \left(\frac{a_{\star}}{R_g}\right)^{7/9} \left(\frac{[1-e_{\star}^2]^2}{1-e_c^2}\right)^{2/9} \ ,
\end{equation}
where $R_g = Gm/c^2$
is the effective gravitational radius of the inner binary.

Introducing the specific orbital parameters for the star S0-2: $a_\star = 1020$~au, $e_\star = 0.88$, $P_{\star} = 15.8$ yr, and assuming $4 \times 10^6 M_\odot$ for the mass of Sgr A*, we estimate the timescale for variations of relevant orbital quantities as a function of the companion's semimajor axis $a_c$.  These estimates are displayed in Fig.\ \ref{fig:timescales}, assuming $10^4 M_\odot$ for the companion mass.  The curve labeled $t_{iEKL}$ arises from the quadrupolar Eqs.\ (\ref{eq:quadrupole}).  Figure \ref{fig:timescales} also shows the GR pericenter precession timescales of both inner and outer orbits from Eqs. (\ref{eq:grprecessin}) and (\ref{eq:grprecessout}), along with the two orbital periods. The point where the $t_{iEKL}$ curve intersects the curve for GR precession of the inner binary [Eq. (\ref{eq:limit})] demarcates the boundary between the colored region where iEKL excitations dominate and instabilities can occur, and the white region, where GR precessions can stabilize the orbits.
 
In addition, we must assume that the binary survives gravitational-radiation decay long enough to be observationally relevant.  One timescale might be 100 years, corresponding (roughly) to the length of an observational campaign.  A more reasonable timescale might be tens of megayears, corresponding to the last star formation episode \citep[e.g.,][]{Lu+13}.  The latter timescale is plotted as the curve $t_{GW}$ in Fig.\ \ref{fig:timescales}.

In the right panel of Fig.\ \ref{fig:allow} displaying $a_c$ vs. $m_c$,
we plot curves indicating where the timescale for iEKL oscillations is equal to (or 10 times) the timescale for the GR pericenter precession of the inner orbit, and where the GW timescale is 100 years and 10 megayears.  The white region then corresponds to companions that we wish to study.

 \subsection{The time variability of a stellar orbit }\label{sec:S02dynamics}
 
Here we focus on potential signatures of the presence of a SMBH binary on 
the orbit of the star S0-2. At the quadrupole level of approximation, the inclination may oscillate and the longitude of ascending node and the pericenter will either librate or circulate, depending on whether the inclination crosses from below to above $90^{\rm o}$.  
From Eq.\ \ref{eq:quadrupole} we can make a rough estimate of the rate of change of the orbital orientation parameters $\Omega_\star$, $\varpi_\star$ and $\theta = \cos i_\star$, of order
\begin{align}
 {\rm Rate} &\sim \frac{3\pi}{4} P_{\star}^{-1} \frac{\eta \alpha^2}{(1-e_\star^2)^2} 
 \nonumber \\
 &\sim 0.004 \, {\rm deg \, yr}^{-1} \left (\frac{m_c}{10^4 M_\odot} \right ) \left ( \frac{a_c}{100 \, {\rm au}} \right )^2   \,.  
\end{align}

The recent closest approach of  S0-2 has been  used to test and confirm the prediction of general relativity for the relativistic redshift  \citep[e.g.,][]{GRAVITY18,Do+19}. This well-studied star provides an opportunity to place limits on the inner orbit's configuration.  The current estimate of the angle of nodes of S0-2 on the sky, $\Omega_{\rm sky}$, is $227^{\rm o}.49 \pm  0.29$(stat) $\pm0.11$(syst) \citep{Do+19} or $228^{\rm o}.075 \pm 0.04$ \citep{GRAVITY18}. 
In addition, the publicly available data from \cite{Do+19} has made it possible to estimate upper limits on a linear drift for each of S0-2's orbital elements using the Keck radial velocities reported in \citet{chu:2018aa} and \citet{Do+19}, the VLT radial velocities reported in \cite{gillessen:2017aa} and the Keck astrometric measurements reported in \cite{Do+19} and expressed within the reference frame developed in \cite{sakai:2019ab} and \cite{jia:2019aa}. The orbital fit methodology is thoroughly described in the Supplementary Materials of \cite{Do+19}. The parameters included in the orbital fit are: the SMBH mass, the distance $R_0$ to the galactic center, the SMBH 2-D position and velocity in the plane of the sky, the SMBH velocity along the line-of-sight, the 6 standard orbital parameters for S0-2, an offset for the  Keck near-infrared imager, NIRC2, radial velocities to correct for fringing effects and two parameters characterizing the correlation within S0-2 astrometric measurements \citep[see][for more details]{Do+19}. In addition, a linear drift for each orbital parameter is included as well (for this analysis, each drift is considered independently from the others). Statistical tests for model selection based on Bayesian evidence \citep[see]{Do+19} show that models that include a linear drift are not favored, such that no significant deviations from zero were reported. An estimate of the 95 \% upper limit on a linear drift of S0-2's orbital elements has been derived from the posterior probability distribution of the fit combined with an estimate of the systematic uncertainty derived from a jackknife analysis at the level of the reference frame construction \citep[see][]{boehle:2016wu,sakai:2019ab,Do+19}. As a result, an upper limit on the rate of change of $\Omega_{\rm sky}$ is estimated as $\left| d\Omega_{\rm sky} /dt\right|< 0.07$~deg~yr$^{-1}$ at the 95\% confidence level; a similar upper limit can be imposed on $|d \omega_{\rm sky} /dt|$  \citep[see also][]{hees:2017aa}.  Further, an upper limit on the rate of change of inclination is estimated as $\left|di_{\rm sky} /dt\right|< 0.02$~deg~yr$^{-1}$. 

Using these constraints we derive potential observational constraints  on the allowable mass and separation of the companion.  
The angle of nodes, orbital inclination and pericenter angle in a given reference basis, either that of the invariable plane or that of the sky, are defined by S0-2's angular momentum unit vector ${\bf j}=(\sin i \sin \Omega, - \sin i \cos\Omega, \cos i)$, and by its Runge-Lenz vector ${\bf e}_{RL}$, leading to a complicated relationship between the orbital elements in the two bases.   However, it is straightforward to show that the quantities
\begin{align}\label{eq:djdt}
\biggl | \frac{d{\bf j}}{dt} \biggr |^2 &= \left (\frac{d i}{dt} \right )^2
  + \sin^2 i \left (\frac{d\Omega}{dt} \right )^2 \,,
\\
\label{eq:eRLdot}
\frac{d{\bf e}_{RL}}{dt} \cdot ({\bf e}_h \times {\bf e}_{RL} ) &= \frac{d\omega}{dt} + \cos i \frac{d\Omega}{dt} \equiv \frac{d\varpi}{dt} \,,
\end{align}
are rotationally invariant, in other words, they have the same value in the basis of the invariable plane as on the sky. Thus, using the aforementioned rate estimates, and adopting  $i_{\rm sky}=133^\circ$ \citep{Do+19}, we obtain upper limits for $|d{\bf j}/dt|_{\rm sky}$ and for $|d\varpi/dt_{\rm sky}|$. On the other hand, combining Eqs. (\ref{eq:bigOmega}) - (\ref{eq:varpi}), we can write:
\begin{eqnarray}
\label{eq:djdtInv}
\biggl | \frac{d{\bf j}}{dt} \biggr |_{\star} &=& \frac{3\pi}{4} \eta \left ( \frac{a_c}{a_\star} \right )^2 \frac{\sin i_\star}{(1- e_\star^2)^{2}} {\cal R}/P_\star \ , \\
\frac{d\varpi}{dt} \biggr |_{\star} &=& \frac{6 \pi G m_\star}{c^2 a_\star (1-e_\star^2)P_\star}+\frac{3\pi \eta}{8}\left(\frac{a_c}{a_\star}\right)^2\frac{{\cal S}P_\star^{-1}}{(1-e_\star^2)^2} \,,
\label{eq:dpomdtInv}\end{eqnarray}
where ${\cal R} = [Q^2 \cos^2 i_\star  + 25 e_c^4 \sin^2 2 \omega_c ]^{1/2}$ and ${\cal S}=4+6e_c^2-3 Q \sin^2i_\star $. Note that the first term in Eq.~(\ref{eq:dpomdtInv}) comes from the GR precession of the star. Since the left hand sides of these two equations are equal to the estimates obtained from data on the sky, we can obtain bounds for $a_c$ as a function of $m_c$.  We show representative solutions in Figure \ref{fig:allow}, where we  depict the $m_c$-$a_c$ parameter space for possible orbital configurations of a SMBH companion. Note that for the inner binary, we have used the relation $\omega_c = \varpi_c - \Omega_\star - \pi$ (Section\ \ref{sec:analytical}), and have chosen without loss of generality   $\varpi_c=0$ \citep[see][]{Naoz+17}. The example bounds shown for $|d{\bf j}/dt|$ are for nominal values of $i_\star = 45^{\rm o}$ and $\Omega_\star= 45^{\rm o}$, and $e_c=0$ and $e_c=0.9$ (solid and dashed lines, respectively). For $d\varpi/dt$ we adopt $i_\star=0^\circ$ (because we are dealing with upper limits from the data, we construct the square root of the sum of the squares of the terms in Eq.\ (\ref{eq:eRLdot})).  This shows that the unknown value of $i_\star$ does not prevent us from placing meaningful constraints on the parameter space, because we have complementary invariant parameters.  For example, it is simple to show that, for a $10^5 \, M_\odot$ companion in a circular orbit, the combined bounds lead to the firm constraint $a_c < 170$~au, irrespective of the mutual inclination of SO-2.  We can generalize this case to the  allowed parameter space as a function of mass:  
\begin{equation}
    a_{c,\rm allowed}(e_c\to0)\lsim 170~{\rm au}\sqrt{\frac{10^5{\rm M}_\odot}{m_c}} \,.
\end{equation}
Conversely, if non-zero values for $d\Omega_{\rm sky} /dt$, $di_{\rm sky} /dt$ and $d\omega_{\rm sky} /dt$ should be obtained, then in the case of $e_c\to0$, Eqs.\ (\ref{eq:djdtInv}) and (\ref{eq:dpomdtInv}) could be used to solve for $i_\star$, independently of the mass and separation of the binary. In other words, the information obtainable on a companion is not strongly dependent on $i_\star$, as shown already in Fig.\ \ref{fig:allow}. 
 The bounds we derive here are more constraining  than the bounds on a companion inferred from data on the proper motion of Sgr A* relative to distant quasars \citep[e.g.,][]{Gualandris+09}. 

We note that at the quadrupole level of approximation the outer orbit's eccentricity remains constant \citep{Naoz+17}.  Since we mostly consider the parameter space in which the GR precession rate is faster than the rate of quadrupole, and thus of octupole, effects,  
we do not expect significant changes in S0-2's eccentricity. The agreement between the right and left panels in Figure \ref{fig:allow}, suggest that octupole effects may be indeed suppressed. S0-2's eccentricity is consistent with the observed upper limit $\left|de/dt\right|< 2.9 \times 10^{-4}$~yr$^{-1}$ at 95 \% confidence level, estimated using the same procedure as described above. Similarly in this regime, the dominant precession of the argument of pericenter of S0-2 (Eq.\ \ref{eq:grprecessout}) should be GR.  But the level of agreement between measured values of the pericenter precession and the GR prediction will still provide bounds on $d\varpi_\star/dt$.  The implications for bounding a companion once the GR precession is actually measured will be explored in future work.

 \subsection{Variability due to interaction with the surrounding medium}\label{sec:IRvir}
 If the mass of the SMBH at the center of the galaxy grew  continuously over the lifetime of the galaxy ($\sim 10^{10}$~yrs), it would imply an average accretion rate of about $4\times 10^{-4}$~M$_\odot$~yr$^{-1}$. However, observational estimates from linear polarization emission from Sgr A*, suggested a current accretion rate of  $10^{-9}-10^{-7}$~M$_\odot$~yr$^{-1}$ \citep[e.g.,][]{Bower+03}.  Roughly speaking, accretion flows onto black holes can be divided into two broad classes: cold, high mass accretion rate, or hot, low mass accretion rate \citep[e.g.,][]{Yuan+14}. The popular model for the accretion disk around Sgr A* is that of a hot, geometrically thick disk which cannot cool efficiently, known as an advection dominated disk \citep{Phinney+81,Narayan+95,Narayan+98,Quataert+02,Quataert+04,Ressler+18}. This accretion model seems to be consistent with the observed spectrum of Sgr A* and may yield high temperature for  the plasma,  \citep[$\sim 10^{12}$~K and $\sim 10^9$~K for the ions and electrons, respectively, e.g.,][]{Narayan+98}.
 Recently, \citet{Murchikova+19} reported the detection of a cool ($\sim 10^4$~K) rotationally supported disk, embedded within the hot plasma. Below we adopt a two component model in order to investigate the consequences of a companion on SgrA*  under a broad range of external conditions.
 
 One of the striking observational features that is associated with accretion flow in the center of the galaxy is the variability in radio, near-infrared (NIR), and X-ray  radiation \citep[e.g.,][]{Zhao+01,Hornstein+02,Ghez+04,Miyazaki+04,Uttley+05,Gillessen+06,Yusef+07,Yusef+11,Neilsen+13,Subroweit+17,Witzel+18,chen:2019aa,Do+19vir}.  The variability can take place on a timescale of minutes to hundreds of minutes, with characteristic coherence timescale of about $243^{+82}_{-57}$~minutes \citep{Witzel+18}.

\begin{figure}
  \begin{center}
    \includegraphics[width=\linewidth]{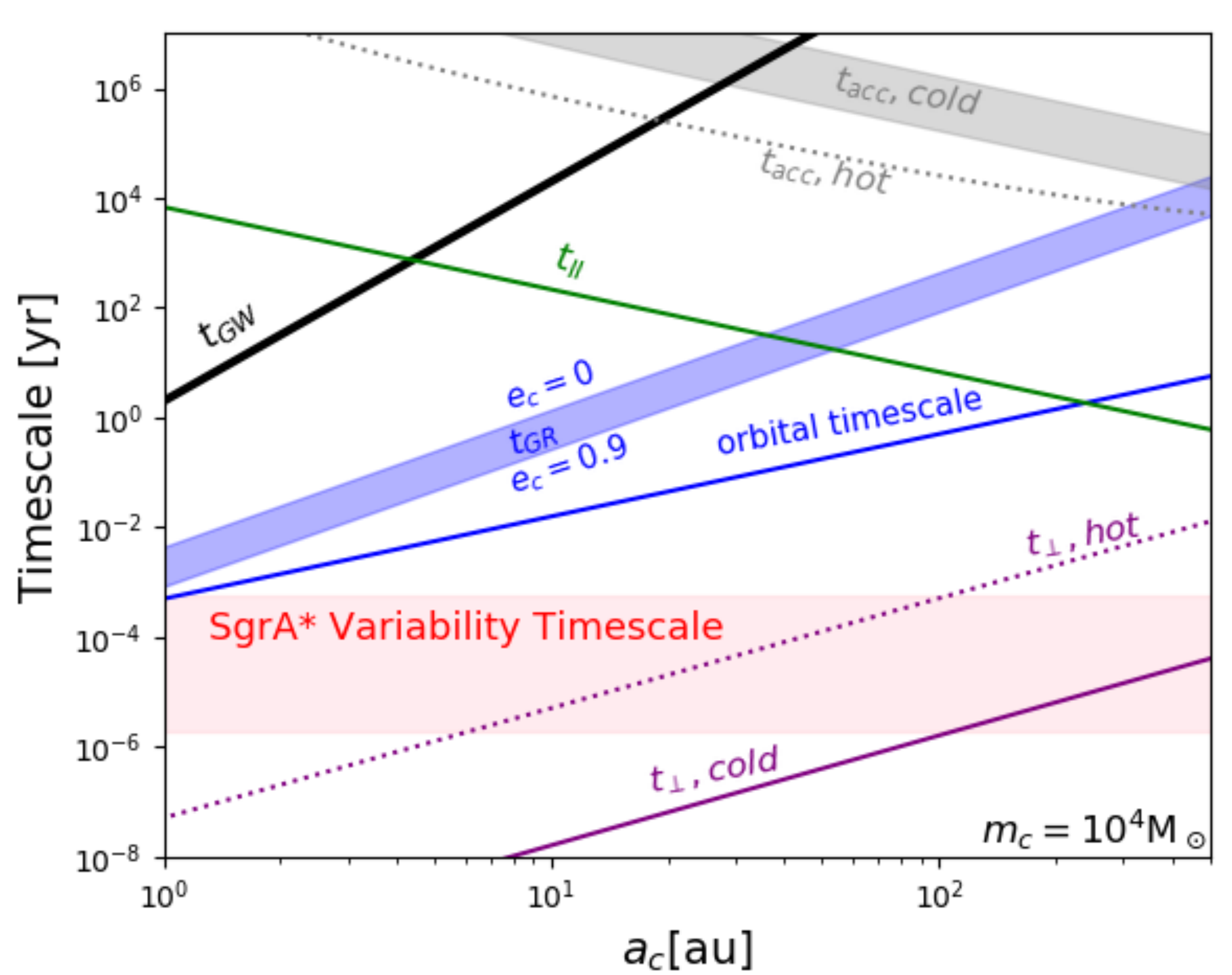}
  \end{center} \vspace{-0.2cm}
  \caption{  \upshape Relevant timescales for interaction between a $10^4$~M$_\odot$ companion and an accretion disk. We consider the following physical processes: Bondi accretion timescales (Eq.~\ref{eq:tacc}], for a hot (dotted grey line) and cold (grey band), disks. The cold disk has a band that represents the uncertainty in the mass estimation of the disk. We also depict the gravitational wave decay timescale (black line, Eq.~\ref{eq:grdamping}), and GR precession for zero and $0.9$ eccentricity (purple band, Eq.~\ref{eq:grprecessin}), as well as the orbital timescale of the companion. We also show the possible type-II migration of a companion in a cold disk (green line, Eq.~\ref{eq:typeII}). Finally we show the minimum timescale associated with a companion crossing the disk perpendicularlly (Eq.~\ref{eq:tperp}), while assuming $e_c=0.9$. We adopt again a cold (purple, solid) and a hot (purple, dashed) disk. It is interesting to point out that this timescale is consistent with Sgr A* variability timescales.      }\label{fig:acc} \vspace{0.2cm}
\end{figure}

 There are several timescales related to a possible companion to the SMBH that may force variability on the emitted radiation. We depict the relevant ones in Fig.~\ref{fig:acc}. The first is associated with the orbital timescale  of the binary companion, which represent the maximal timescale for perturbation of the disk due to orbital crossing. It has negligible dependency on the companion mass (because $m_c<m_\bullet$). 
 The binary orbit may not lie in the plane of the disk, and thus may only interact with it when the two cross. We can estimate the minimal time due to a perpendicular configuration of the orbit relative to the disk. In that case the orbit interacts with and perturbs the disk for 
  \begin{equation}\label{eq:tperp}
t_{\perp}\sim \frac{H}{v_{K,peri}} \ ,
\end{equation}
where $v_{K,peri}$ is the Keplerian velocity of the companion at pericenter, and $H$, is the scale height of the disk at the point of interacting with the companion, estimated as:
\begin{equation}
H= \frac{c_s}{\Omega_{K,peri}} \ ,
\end{equation} 
where $\Omega_{K,peri}$ is the Keplerian orbital frequency of the disk around Sgr A* at the point of the interaction with the companion. The speed of sound, $c_s$ can be estimated for ideal gas with adiabatic index, $\gamma$, and temperature $T$ by 
   \begin{equation}
c_s= \sqrt{\frac{2k_B\gamma T}{m_p}} \ ,
\end{equation}
where $m_p$ is the mass of a proton and $k_B$ is the Boltzmann constant. In Figure \ref{fig:acc} we explore two possible scale heights, one of the cold disk (solid lines) and one for the hot component (dotted lines); we assumed an eccentric $10^4$~M$_\odot$ companion with $e_c=0.9$.
In Figure \ref{fig:acc} we show $t_{\perp}$ for the nominal system, where interaction of a companion perpendicular to both hot and cold components may perturb the disk on timescales that are consistent with observations. 

The companion may also accrete mass as it travels within the disk. Assuming Bondi accretion, the mass accretion rate onto the companion black hole is
   \begin{equation}
\dot{M}_c\sim \frac{4\pi G^2 m_c^2 \rho}{(c_s^2+\langle v_{K,r}\rangle^2)^{3/2}}  \ ,
\end{equation}
where $\langle v_{K,r}\rangle$ is the average companion's Keplerian velocity along the orbit relative to the Keplerian velocity of the disk, and  $\rho$ is the density of the disk, which can be estimated from the number density $n$ of the disk. The latter is estimated as $130$~cm$^{-3}$ for the hot component \citep{Baganoff+03} and as $10^5$~cm$^{-3}$ for the cold disk \citep{Murchikova+19}. The timescale associated with accretion is then
   \begin{equation}\label{eq:tacc}
t_{acc}\sim \frac{M_d}{\dot{M}_c}  \ ,
\end{equation}
where $M_d$ is the mass of the disk. Thus, for the cold disk, with $M_d\sim 10^{-4}-10^{-5}$~M$_\odot$ \citep{Murchikova+19} the timescale due to accretion is rather long, depicted in Figure \ref{fig:acc} as a grey band. For the hot component we adopt $M_d\sim 10^{-9}$~M$_\odot$.

As can be seen from Figure \ref{fig:acc}, the crossing time of an eccentric companion, perpendicular to the disk (Eq.~\ref{eq:tperp}), is consistent with the IR variability timescale of $\sim 1-300$ minutes. We note that we do not expect the effect to be periodic as the disk may precess, warp and torque because of the companion. Moreover the main variability driver may be due to the gas rearranging itself which is not necessarily  periodic. The complex physical processes that should be taken into account are beyond this  proof-of-concept calculation. We also do not expect a periodic effect because the accretion is very low, \citep[unlike OJ 287, that may have a large accreation rate and a dense disk ][]{Sillanpaa+88,Lehto+96,Valtonen+19}. 

We estimate the change of luminosity that occurs when the companion passes through the disk by considering the ratio between the Bondi accretion surface of the companion $\sim \pi r_B^2$, where $r_B\sim G m_c/c_s^2$, to the surface area of the annulus in the disk $\pi a_c r_B$:
  \begin{equation}
\bigg|\frac{\Delta L}{L}\bigg| \lsim \frac{G m_c}{a_c c_s^2} \sim 3\% \left(\frac{m_c}{10~{\rm M}_\odot}\right) \left(\frac{10~{\rm au}}{a_c} \right)\left( \frac{10^9~{\rm K}}{T}\right)\ .
\end{equation}
This is the minimum luminosity difference we expect from the companion plunging through the disk. The rearranging of the gas following the perturbation to the disk is expected to reach larger amplitudes. We speculate that it may even reach the observed variability magnitude $\sim 10-75\%$
 \citep{Witzel+18}. We  caution, however, that a hydrodynamics simulation is needed to explore the full potential of this model in addressing the variability of Sgr A*.

Finally we note that if the companion lies within the plane of a cold disk, it may open a gap and migrate inwards via type II migration with the estimated timescale of \citep[e.g.,][]{Armitage07}
  \begin{equation}\label{eq:typeII}
t_{II}\sim \frac{2}{3\alpha}\left(\frac{H}{a_c}\right)^{-2} \Omega_{K,peri}^{-1}\ ,
\end{equation}
where a gap opening condition is possible for a companion mass as low as $\sim 10$~M$_\odot$, inward to $\sim 500$~au. Thus, in the presence of a cold disk, the companion could migrate inward on a faster timescale than gravitational wave decay. Moreover, this process may result in spiral arms that will result in possible observed signatures of their on.  
 We note that since the gas densities in the vicinity of the SMBH are small, the the gas drag on the binary timescales, are much longer than $t_{\rm GW}$,  \citep[see][Eq. (52)]{Antoni+19}.

\begin{figure}
  \begin{center}
    \includegraphics[width=1\linewidth]{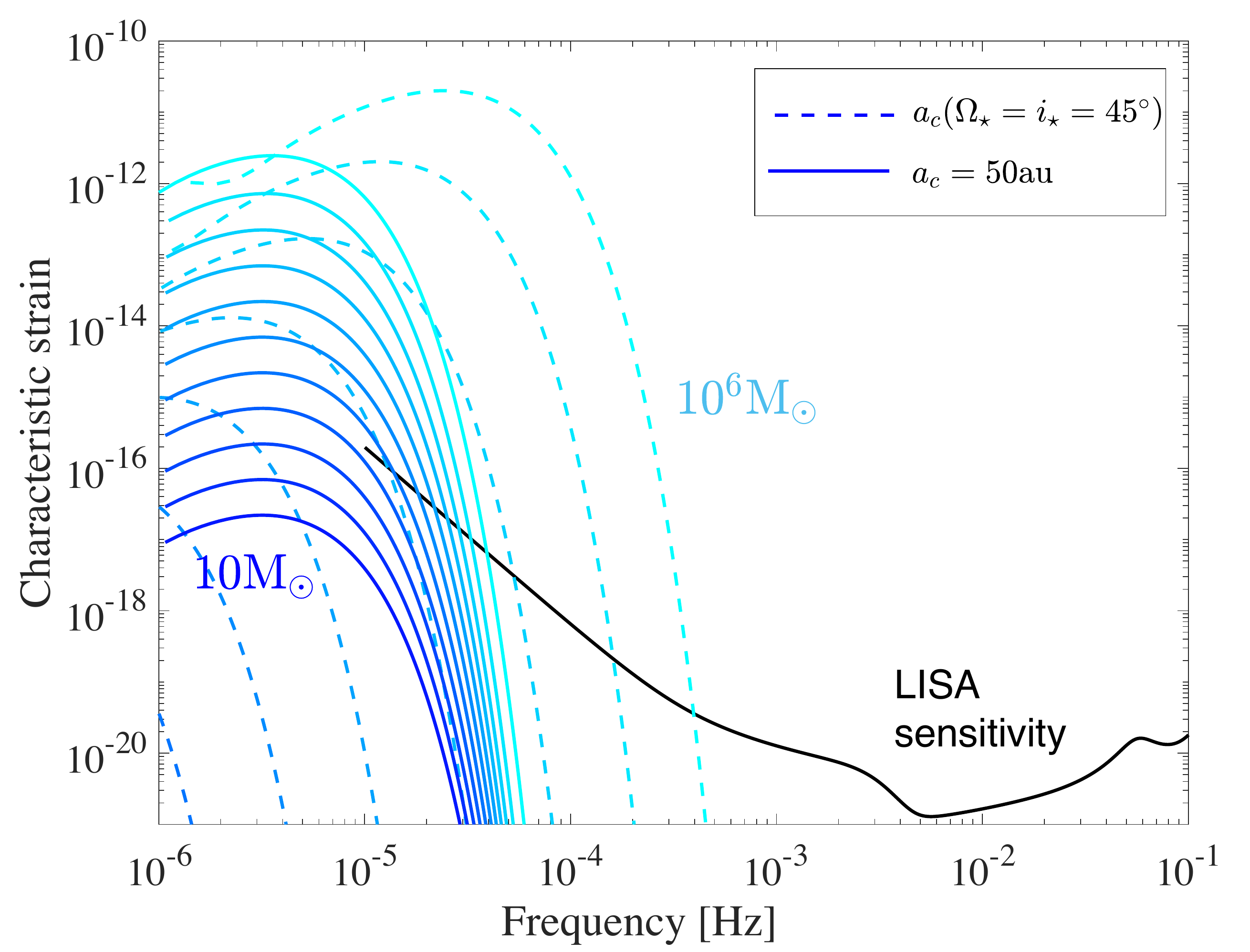}
  \end{center} 
  \caption{  \upshape Characteristic strain as a function of frequency. We consider the gravitational wave signal from a SMBH binary located at $a_=50$~au (solid lines) as well as $a_c(\Omega_\star=i_\star=45^{\circ})$, which corresponds to the red dashed line from the left panel in Figure \ref{fig:allow}. 
   We consider a range of masses varying from $10$~M$_\odot$ (dark blue)  to   $10^6$~M$_\odot$ (light blue). In all cases we adopt $e_c=0.9$. We adopt a LISA observational time of $4$~yrs. The LISA noise sensitivity is shown in  black \citep{Robson+18}.  }\label{fig:GW} \vspace{0.2cm}
\end{figure}
 
 \subsection{Gravitational Wave signal}\label{sec:GW}
 
 A massive companion in orbit around Sgr A* will emit gravitational waves.  Since the companion can be on an eccentric orbit,  
the GWs are emitted over a wide range of frequencies that approximately peaks at a frequency of \begin{equation} f_p(a_c,e_c) = (1 + e_c)^{1/2}(1 - e_c)^{-3/2} f_{\rm orb}(a_c) \ , \end{equation} where $f_{\rm orb}(a_c) = (2 \pi)^{-1}\sqrt{G(m_{\bullet} + m_c)} a_c^{-3/2}$.
To quantify the parameter space where a companion is detectable in the LISA band, we estimate the signal-to-noise ratio (SNR) as a function of $a_c$ and $e_c$ as: \citep[e.g.][]{Robson+18}:
\begin{equation}\label{eq:SNR}
{{\rm SNR}^2(a_c,e_c)} = \int \frac{h^2_c(a_c,e_c,f)}{f^2 S_n(f)} df \ , \vspace{0.1cm}
\end{equation}
where $S_n(f)$ is the effective noise power spectral density of the detector, weighted by the sky and polarization-averaged signal response function of the instrument  \citep[e.g., Eq.~(1) in][]{Robson+18}. Using the equations presented in \citet{Kocsis+12} and \citet{Hoang+19} we calculate the characteristic strain $h_c(a_c,e_c,f)$ from the GW radiation as a function of the frequency. As can be seen in Figure \ref{fig:GW}, a massive companion may be detectable by LISA for a non-negligible part of the parameter space. 

\section{Discussion} 
The hierarchical nature of galaxy formation suggests that a SMBH binary could exist in our galactic center. In this paper we have proposed ways to constrain the possible orbital configuration of such a binary companion to Sgr A*. In particular we focused on the well studied star S0-2 and showed that requiring its stability in the presence of a companion to Sgr A* yields interesting constraints on the possible allowed configurations of such a companion (Figure \ref{fig:allow}, right panel). We then pointed out that measurements of the time variations in the orbital parameters of S0-2 yield much stronger constraints (Figure \ref{fig:allow}, left panel) and that improved observations  could even lead to the detection of a companion to Sgr A*.  We note that expanding this exercise to other stars at the galactic center is straightforward, and could yield tighter or complementary constraints. In particular, precise measurements of the time variability of the orbital parameters for other stars will allow narrowing the parameter space.    

 We note that a companion to Sgr A* may also result in an imprint of the ejection velocity distribution of hypervelocity stars \citep[e.g.,][possibly detectable by Gaia]{Marchetti+18,Rasskazov+19}. Hypervelocity stars are thought to be generated from the unbinding of binary stars approaching too close to an SMBH \citep{Hills88}. Stellar binaries unbinding due to gravitational interaction with binary SMBH can result in extreme velocities for the ejected  stars, potentially providing a unique signature for the existence of this massive binary \citep[e.g.,][]{Darbha+19,Rasskazov+19}. 

A SMBH companion could also interact with the accretion disk at the galactic center.   As a proof-of-concept, we derive timescales estimates for the effect of a companion on 
the surrounding disk. We found consistency between the observed order of magnitude IR variability  and a companion that plunges into a disk.   
Finally, we showed that a companion to Sgr A* could be observable via the space gravitational-wave detector LISA.

\acknowledgements
We thank the anonymous referee for their report. SN and CMW thank the UCLA Bhaumik Institute for Theoretical Physics for the hospitality that enabled the initiation of this project. SN acknowledges the partial support of NASA grant No. 80NSSC19K0321, and also thanks Howard and Astrid Preston for their generous support.  CMW is supported in part by the National Science Foundation under Grant Nos. PHY 16-00188 and 19-09247. ERR is indebted to the the Heising-Simons Foundation and the Danish National Research Foundation (DNRF132) for support. TD is funded by NSF AAG Grant 1909554.  SN, AMG and TD thank the Keck foundation for its partial support of the NStarsOrbits Project

\begin{figure*}[t]
     \centering
         \includegraphics[width=0.5\linewidth]{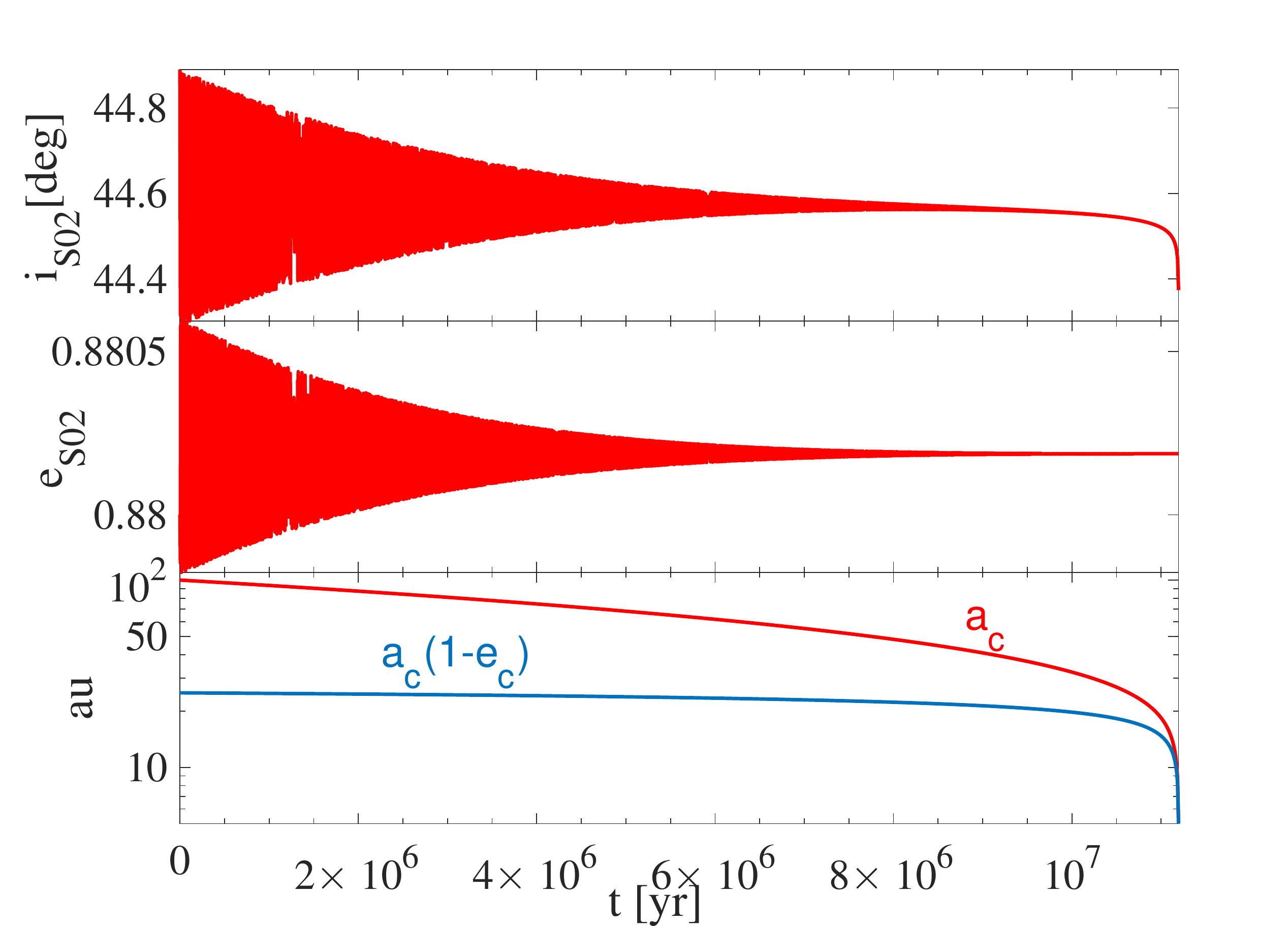}
          \includegraphics[width=0.5\linewidth]{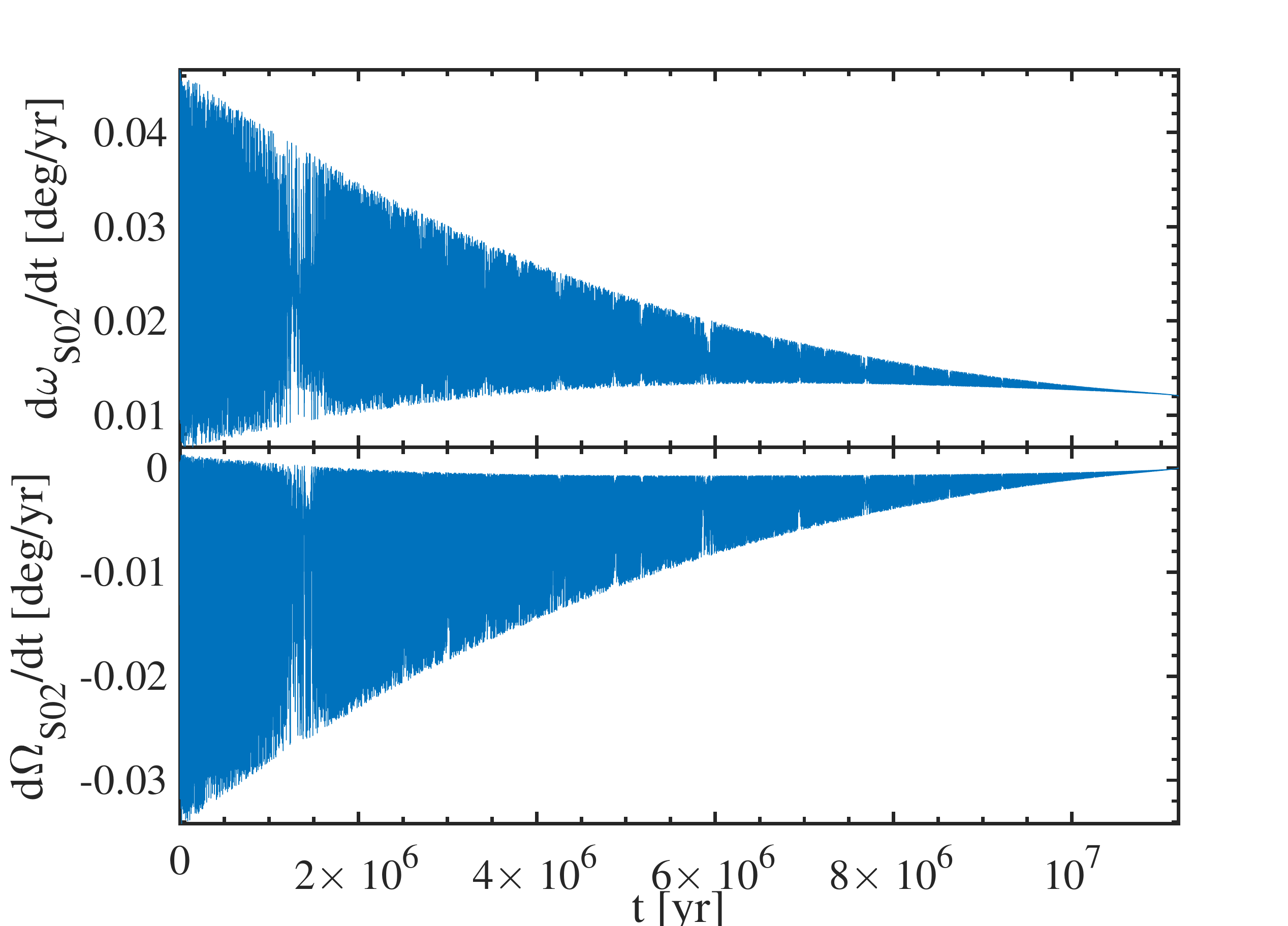}
      \caption{ \upshape  Numerical evolution of a representative system of S0-2 orbiting an inner binary with $m_c=10^4$~M$_\odot$, including both quadrupole and octupole perturbations. We set the initial values $a_c=100$~au, $\Omega_\star=i_\star=45^{\rm o}$, $e_c = 0.9$ and $e_\star = 0.88$. Note the effect of the octupole-level of approximation \citep{Naoz+11sec} which results in negative $d\Omega/dt$. }
     \label{fig:example}
 \end{figure*}
 
\appendix 

\section{Numerical Example}\label{sec:numeric}
 To illustrate the long-term evolution of system consisting of an inner SMBH binary system and an outer star such as S0-2, we choose a nominal system for which $m_c=10^4$~M$_\odot$ with initial eccentricity of $0.8$ and with a mutual inclination with S0-2 of  $45^\circ$. We include quadrupole and octupole perturbations as well as GR pericenter precessions, and integrate the secular orbit element equations over 11 megayears.  As can be seen in Fig.\ \ref{fig:example}, not only are the outer orbit's eccentricity excitations of small amplitude but also as the BH binary orbit shrinks and circularizes, the oscillations damp out. However, the rate of change of $\Omega_{\star}$ and $\omega_{\star}$ remain significant over a substantial fraction of the evolution.


\bibliographystyle{hapj}

\end{document}